\documentstyle[epsfig,longtable]{aipproc}

\begin{document}

\title{X-ray Observations of AGN at Intermediate to High Redshift}
 
\author{K. A. Weaver$^*$}
\address{$^*$NASA/GSFC, Code 662, Greenbelt, MD 20771}

\maketitle

\begin{abstract}

The cores of active galactic nuclei (AGN) harbor some of the most 
extreme conditions of matter and energy in the Universe. 
One of the major goals of high-energy astrophysics is to 
probe these extreme environments in the vicinity of 
supermassive black holes, which are intimately linked to the
mechanisms that produce the continuum emission in AGN.  X-ray studies 
seek to understand the physics responsible for the continuum emission,
its point of origin, how nuclear activity is fueled,
and how supermassive black holes evolve.  
The key to finding answers to these questions 
lies in measuring the intrinsic luminosities and spectral shapes,
the relation of these properties to other wavebands,
and how the source properties 
change with redshift.  This article reviews X-ray
observations of AGN from redshifts of $\sim0.1-3$ with the
goal of summarizing our current knowledge 
of their X-ray spectral characteristics.  
Results are evaluated in terms of their robustness and are examined 
in the light of current theoretical predictions of energy release via
processes associated with the accretion mechanism. 
A possible evolutionary scenario is discussed,
along with the importance of AGN studies at high 
redshift as they relate to 
the total energetics of the Universe.

\end{abstract}

\section*{Introduction}

Since their discovery, active galactic nuclei (AGN) have stood out as
uniquely luminous objects in the Universe.  We are fairly confident 
that their ultimate power source is the release of gravitational energy
sustained by an accretion disk, which feeds matter directly onto a
supermassive black hole.  Evidence for the accretion mechanism is 
found in X-ray-bright Seyfert galaxies, which have broad Fe K$\alpha$ lines
indicative of radiatively efficient, geometrically thin accretion disks
extending down to the radius of marginal stability\cite{nandra97a,tanaka95}.
However, the mechanisms to produce electromagnetic radiation from
the accreting material and the manner in which this material
actually reaches the black hole are still unclear.  Understanding the 
accretion mechanism is important because nuclear activity 
in galaxies is common (perhaps more common than previously thought) and
therefore accretion is likely to play a fundamental role in the 
energetics of the Universe.

The Universe is populated with many classes of AGN,
the most powerful of which are 
QSOs ($L_X \sim 10^{46-48}$ erg s$^{-1}$).
Their high luminosities make them observable at great distances,
thereby providing ways to obtain 
fundamental information about the formation and
subsequent evolution of galaxies.  As we look toward the early Universe,
studies of QSOs can help answer these
important questions:\hfill\break

$\bullet$ Do all galaxies contain massive black holes and  
what roles do AGN play in the formation and evolution
of galaxies?\hfill\break

$\bullet$ How does the accreting material
make its way into the surroundings of the black hole,
and how is this material fed directly into the black hole?\hfill\break

$\bullet$ Is there a change
in the accretion efficiency or accretion rate with z?\hfill\break

$\bullet$ Are we seeing the flaring of short lived QSO events in
many nuclei or a slow decline in a few nuclei that have been
QSOs from the start?\hfill\break

Or, in observational terms:
{\it In what ways do AGN exhibit spectral evolution?}\hfill\break

X-ray observations are of particular value because they provide 
a powerful diagnostic of the environs of the accretion flow and  
a powerful means for tracing evolution.  Variability studies show 
that the X-ray continuum emission in AGN originates on the spatial 
scales we are most interested in --   
close to the black hole.  X-rays can also 
penetrate large amounts of gas and
dust in which some active nuclei are embedded.  Moreover,
X-ray emission appears to be a universal property of QSOs\cite{at86},
which allows us to trace their properties out to high redshift.  
The major limitation of X-ray studies is the need for   
different instruments to cover the entire X-ray continuum range.
Unbiased and statistically valid X-ray samples of QSOs have been 
difficult to obtain, but this will change with the 
next generation of X-ray observatories, 
beginning with the launch of $\it Chandra$ and $XMM$ in 1999.
This article evaluates the current status of X-ray spectroscopy studies of
QSOs, with emphasis on objects whose spectra appear to be dominated 
by accretion mechanisms rather than jet/beaming mechanisms.

\section*{X-ray Continuum Properties of QSOs}

\begin{table}[b!]
\caption{The X-ray Spectral Properties of QSOs}
\label{table1}
\begin{tabular}{lccccccrr}
Radio & Sample& z$^a$ &Energy& Instr.&$<\Gamma_{\rm X}>$& Intrinsic$^b$ 
&Comments$^c$&Ref.\\
class & size &  & [keV] &  &  & N$_{\rm H}$? & & \\        
\tableline
RQQ &42 &$0.12\pm0.05$& $0.1-2.4$ & $Rosat$  &$2.56^{+0.10}_{-0.11}$& 
no & $^d$ & \cite{swft96}\\
  & 9 &$0.3\pm0.03$ & $0.1-2.4$ & $Rosat$  &$2.47\pm0.33$   & no & 
$^d$ & \cite{swft96} \\
  &19 &$0.19\pm0.08$& $0.1-2.4$ & $Rosat$  &$2.72\pm0.09$   & no & 
    $^d$ & \cite{l97} \\
 &390&$0\rightarrow2.5$ & $0.1-2.4$ & $Rosat$ &$2.58\rightarrow2.22$  
     & ---& $^d$ & \cite{ybsv98}\\ 
    &16 &$0.18\pm0.21$& $0.3-3.5$ &$Einstein$&$1.91^{+0.67}_{-0.36}$& 
no & s. excess (7) & \cite{we87}\\ 
   &12 &$0.076\pm0.04$& $0.1-10$  &$Exosat$  &$2.18\pm0.35$   & no & s. 
excess (5) & \cite{comastri92}\\
  & 9 &$0.54\pm0.47$&$0.5-10$   &$ASCA$    &$1.93\pm0.06$   & yes 
(3) &Fe K (5) $^d$ & \cite{reeves97} \\ 
  & 5 &$2.1\pm0.13$ & $2-10$    &$ASCA$    &$1.68\pm0.09$   & no & 
FeK (1) $^d$ & \cite{v98} \\
  & 7 &$0.13\pm0.08$& $2-20$    &$Ginga$   &$1.90\pm0.38$   & ---& 
   & \cite{w92} \\ 
RLQ &65 &$0.08\rightarrow2.3$& $0.1-2.4$ &$Rosat$ 
   &$2.52\rightarrow1.87$ & no & $^d$ & \cite{swft96} \\ 
  & 4 &$0.27\pm0.11$& $0.1-2.4$ &$Rosat$   &$2.15\pm0.14$   & no & 
     $^d$ & \cite{l97}\\
  & 4 &$3.16\pm0.23$& $0.1-2.4$ &$Rosat$   &$1.71\pm0.08$   &yes (3) 
& $^d$ & \cite{e94} \\
  & 9 &$2.56\pm0.8$ & $0.1-2.4$ &$Rosat$   &$1.53\pm0.06$   &yes (2) & 
      $^d$ & \cite{c97}\\
    &17 &$0.34\pm0.19$& $0.3-4.5$ &$Einstein$& $1.48^{+0.63}_{-0.36}$ & 
no &  & \cite{we87} \\
   & 5 &$0.27\pm0.22$& $0.1-10$  &$Exosat$  & $1.79\pm0.19$  & no & s. 
excess (1) & \cite{comastri92} \\
  & 3 &$2.3\pm0.83$ & $0.5-10$  &$ASCA$    &$1.67\pm0.20$   &yes 
(1) & $^d$ & \cite{s96} \\ 
  &15 &$2.42\pm1.29$& $0.5-10$  &$ASCA$    &$1.63\pm0.04$   &yes 
(9)&Fe K (2) $^d$ & \cite{reeves97}\\ 
  & 9 &$2.56\pm0.8$ & $0.5-10$  &$ASCA$    &$1.61\pm0.04$   &yes 
(6) & $^d$ & \cite{c97} \\
  & 6 &$0.38\pm0.3$ & $2-20$    &$Ginga$   & $1.71\pm0.16$  & ---& 
     $^d$ & \cite{w92}\\
\end{tabular}
Notes: $^a$Mean redshift and standard deviation except for cases with 
arrows, which indicate the range of z.  $^b$Indicates 
whether absorption in excess of the Galactic value is present.  
Parenthesis contain the number of objects.
$^c$Indicates cases for which excess soft X-ray emission or
Fe K$\alpha$ emission is detected. 
Parenthesis contain the number of objects. 
$^d$Denotes data that are plotted in Figure 1. 
\end{table}

The X-ray spectral characteristics of QSOs
for redshifts up to z $\sim3$ are summarized in Table 1.  All 
results derive from spectral modeling techniques, which assume that the
QSO spectrum in a given energy band can be modeled with an absorbed power
law having a photon index $\Gamma$ ($N(E)\propto$ E$^{-\Gamma}$).  For cases
where the line-of-sight absorption is consistent with that due to 
our Galaxy ($N_{\rm Hgal}$),
the fits have $N_{\rm H}$ fixed at $N_{\rm Hgal}$.  For cases where 
$N_{\rm H}$ is significantly larger than $N_{\rm Hgal}$, $N_{\rm H}$ is 
left as a free parameter.  In this article, the term ``soft X-ray''
is loosely applied to photon energies between 0.1 and 2.4 keV and the 
term ``hard X-ray'' is loosely applied to 
photon energies between 2 and 10 keV.
The energy bands of the experiments listed in Table 1
overlap and so only the 
photon indices that are strongly weighted toward soft or hard energies 
are discussed.   

When photon index is plotted against redshift (Figure~\ref{firstfigure}), 
it is apparent that $\Gamma$ decreases with increasing z. 
For RQQs, $\Gamma$(soft) ranges from  $\sim2.6$ at low z to 
$\sim2.2$ at high z ($\Delta\Gamma=0.4$ from $z=0.1\rightarrow2$) 
while $\Gamma$(hard) changes only slightly with z, from $\sim1.9$ to 
$\sim1.7$.  For RLQs, $\Gamma$(soft) decreases by a larger amount from 
$\sim2.5$ at low z to $\sim1.7$ at high z ($\Delta\Gamma=0.8$
from $z=0.1\rightarrow3$), while $\Gamma$(hard) remains 
fairly constant with z.  In addition, the soft X-ray index 
is related to
radio loudness in the sense that RLQs have systematically
smaller values of $\Gamma$(soft) than RQQs.
At high energies, the spectral shapes are similar  
implying a common emission mechanism and minimal 
spectral evolution.

What about possible selection biases?
The set of observations that most likely contain
a selection bias is the $ASCA$ sample of high-z
RQQs\cite{v98}.
For a given optical luminosity, RQQs are $\sim3$ times less 
luminous in X-rays than RLQs\cite{z81} and so only the most 
luminous RQQs have reliable hard X-ray data, especially at high z.
For a given distribution of $\Gamma$(hard), $ASCA$ may be
biased towards detecting objects with small $\Gamma$(hard). 
On the other hand, these objects still
possess fairly steep soft X-ray slopes\cite{ybsv98}.  Since 
these objects have both steep low-energy spectra {\it and} 
flat high-energy spectra,  
the trend seen with $ASCA$ probably does represent 
RQQs at high z.

The dichotomy of spectral 
indices is robust and provides strong evidence for two distinct 
emission mechanisms, one at low energies
that dominates the spectra up to z $= 1-2$, and one at high 
energies that is approximately independent of z. 
A distinct energy for this spectral
``break'' toward low energies (obviously a larger effect
in RQQs than in RLQs) has not
been found, but it is probably somewhere between 0.5 and 1 keV
for low-z objects\cite{comastri92}.
To explain the spectral changes with z,  
RLQs must have their soft 
component shifted out of the $Rosat$ band by z = 2, beyond
which $\Gamma$ reaches its redshift-independent 
value\cite{bys97}.  RQQs, on the other hand,
have not yet displayed the point at which the soft component is
shifted out of the $Rosat$ band, and so the spectral break must 
occur at a higher energy compared to RLQs (Figure 1).
Depending on how the soft and hard X-ray  
components are normalized, 
either the soft X-ray emission is enhanced 
in RQQs relative to RLQs or the 
hard component is enhanced in RLQs relative to RQQs.
The fact that RLQs are the stronger X-ray sources
implies the latter.

\section*{Photoelectric Absorption}

\begin{table}[b!]
\caption{X-ray Properties of a Representative Sample of QSOs at z $>2$}
\label{table2}
\begin{tabular}{lcccccccc}
Quasar  & z & $\Gamma_x$ & N$_{\rm H}^a$ & N$_{\rm Hgal}$ & log 
$L_x$ 
     & Radio & Fe K$\alpha$ EW$^b$ & Ref. \\
 & & & [$10^{20}$ cm$^{-2}$] & [$10^{20}$ cm$^{-2}$] &[ergs s$^{-1}$]  
& Class & [eV] &  \\
\tableline
S5 $0014+81$  & 3.38  & $1.7\pm0.07$           & $27.4\pm4$         & 
13.9
    & 47.8 & RLQ & $<70$ & \cite{c97}  \\
    & & & $554^{+196}_{-170}$ (qf) & & & & &  \\
$0040+0034$     & 2.00  & $1.79^{+0.15}_{-0.14}$ & $9.7^{+5.5}_{-5.2}$& 
2.45
    & 46.4 & RQQ & --- & \cite{v98}  \\
PKS $0237-233$  & 2.22 & $1.68\pm0.06$           & $122^{+62}_{-54}$ (qf) & 
2.39
    & 46.8 & RLQ & $<32.2$ & \cite{reeves97}  \\
$0300-4342$     & 2.30  & $1.69^{+0.24}_{-0.13}$ & $<6.8$             & 
1.83
    & 46.0 & RQQ & ---  & \cite{v98}  \\
Q $0420-388$     & 3.12  & $2.24\pm0.48$          & $34\pm33$ (qf)          
& 1.91    
    & 46.9 & RLQ & ---   & \cite{e94}   \\
PKS $0438-436$& 2.85  & $1.5\pm0.1$          & $5.8^{+2.6}_{-1.4}$& 
1.47
    & 47.0 & RLQ &$<240$ & \cite{c97} \\
    & & & $72^{+44}_{-22}$ (qf) & & & & &  \\
PKS $0528+134$  & 2.07  & $2.64\pm0.07$          & $420\pm90$ (qf)  & 
23.0
    & 47.1 & RLQ & $119\pm58$ & \cite{reeves97} \\
PKS $0537-286$& 3.11  & $1.4\pm0.1$           &$2.8^{+1.0}_{-0.7}$ & 
1.95
    & 47.3 & RLQ &$<139$ & \cite{c97} \\
    & & & $16.7^{+19.3}_{-14.6}$ (qf) & & & & &  \\
S5 $0836+71^c$  & 2.17  & $\sim1.5$              & $\sim3.3\pm0.7$    & 
2.78
    & 47.5 & RLQ &$<110$ & \cite{c97} \\
$1101-264$      & 2.15  & $2.19^{+0.58}_{-0.47}$ &$19.5^{+19.5}_{-
15.5}$& 5.68
    & 45.8 & RQQ &$690\pm560$ & \cite{v98}  \\
$1255+3536$     & 2.04  & $1.59^{+0.12}_{-0.11}$ & $<6.6$             & 
1.22
    & 46.3 & RQQ & ---  & \cite{v98}  \\
$1352-2242$     & 2.00  & $1.66^{+0.25}_{-0.23}$ & $<17.9$            & 
5.88
    & 46.0 & RQQ & ---  & \cite{v98}  \\
$1422+231$      & 3.62  & $1.68\pm0.14$          & $<151$ (qf)     & 
2.9
    & 47.0 & RLQ   & $<263$    & \cite{reeves97} \\
RXJ $1430.3+4203$&4.72  & $1.29\pm0.05$          & $<5.3$             & 
1.4
    & 47.1 & RLQ & $<100$ & \cite{fabian98} \\
Q $1508+571$      & 4.30 & $1.43\pm0.08$          & $<477$  (qf)     & 
1.34
    & 47.2 & RLQ & $<156$  & \cite{reeves97} \\
PKS $1614+051$   & 3.21 & $1.43\pm0.14$          & $<250$ (qf)    & 
5.0
    & 46.6 & RLQ &$<132$ & \cite{reeves97}  \\
Q $1745+624$     & 3.89 & $1.68\pm0.25$     & $21^{+15}_{-14}$   & 
3.4
    & 47.0 & RLQ & $<180$  & \cite{k97}  \\
    & & & $61^{+108}_{-59}$ (qf) & & & & &  \\
PKS $2126-158$ & 3.28 & $\sim1.6\pm0.1$        & $10.4^{+3.1}_{-2.3}$& 
4.85
    & 48.0 & RLQ &$<107$ & \cite{c97} \\
    & & & $120^{+80}_{-50}$ (qf) & & & & & \\
PKS $2149-306$ & 2.35 & $1.54\pm0.05$            & $8.3^{+1.8}_{-2.2}$& 
1.91
    & 47.8 & RLQ & $<85$ & \cite{c97}  \\
PKS $2351-154^d$ & 2.67 & 1.92(fixed)     & $12.7^{+3.7}_{-
2.9}$& 2.18
    & ---  & RLQ & --- & \cite{schartel97} \\
    & & & $200^{+90}_{-60}$ (qf) & & & & & \\
\end{tabular}
Notes:
$^a$(qf) absorption in the QSO frame
as opposed to the observer's frame (all others).\\
$^b$Assumes a narrow Gaussian line at 6.4 keV in the
QSO rest frame.\\
$^c$Absorption changes by $\Delta$N$_{\rm H}\sim8\times10^{20}$ in 0.8 yr.\\
$^d$Absorption changes on timescale of $<0.41$ yr in the QSO frame.\\
\end{table}

Characteristics for QSOs at z $>2$ that have 
high-quality X-ray data are listed in 
Table 2\footnote{Not necessarily a complete sample.}.  
More than 1/2 of the RLQs possess absorption 
in excess of the Galactic value.\footnote{A strong case
for absorption is made by Fiore et al. (1998) \cite{fiore98}. 
An alternative but less favored explanation 
is downward curvature in
the intrinsic spectrum, such as that resulting from emission
dominated by synchrotron losses.}
In contrast, RQQs and low z quasars lack significant
amounts of intrinsic absorption, with
a handful of RQQs, such as PG 1114+445 \cite{george97}
showing evidence for ionized absorption.  
The absorption in RLQs can be as much as 
$\sim 5 \times10^{22}$ cm$^{-2}$ in the quasar frame,
which is similar to that seen in intermediate-type 
Seyfert galaxies such as NGC 4151.  However, the  
physical properties of the absorbers are not well known
because the X-ray data are ambiguous. 
In some cases, data at other wavebands support
the contention that the absorption is 
physically associated with the quasar\cite{e98}, while 
in other cases, the data favor absorption at low 
z \cite{s94}.  If the absorber is intrinsic to the 
quasar, it could be nuclear material as in low-z, low-$L_x$
objects or it could exist on a
larger scale such as the host galaxy (or protogalaxy).

Because of the uncertainties associated 
with the properties of the absorbing material in 
high-z RLQs, some general spectral trends are not clear.
For example, does the shape of the intrinsic spectrum  
depend on the luminosity of the source?
Other QSO studies have not found 
a significant correlation between $\Gamma$ and L$_X$
except for the general trend that
RLQs have smaller X-ray indices than
RQQs\cite{reeves97,w92,c97}.  Table 2 also  
shows no correlation between $\Gamma$ and L$_X$, but   
$\Gamma$ depends on how the absorption is 
handled in the spectral modeling.  
Having $N_{\rm H}$ ``wrong'' can
make $\Gamma$ artificially large or small and 
so the true values of $\Gamma$ are somewhat uncertain.  

The data also do not require that {\it only} 
high-z RLQs possess intrinsic absorption.
Indeed, if only the hard X-ray component is 
absorbed, this underlying absorption 
could remain ``hidden'' in other QSOs.   In RQQs, the 
stronger soft component might easily swamp the 
underlying absorption, regardless
of z.  For RLQs, where the soft component 
is not as prominent, column densities of
a few $\times10^{22}$ cm$^{-2}$ would flatten the 
observed spectrum at $\sim1-2$ keV in the quasar frame.      
Such flattening could go undetected in low-z quasars in 
the following ways.  For $Rosat$, the 
flattening falls too near the upper energy range 
of the detector.  For $ASCA$, which covers a larger 
bandpass, flux from the soft component 
may average out the flattening so that it is not detected.
On the other hand, for z = $1-2$ the 
contaminating soft component is
shifted out of the $ASCA/Rosat$ band and 
the absorption cutoff is shifted to around 
$\sim0.3-0.6$ keV in the observer's frame, into a region 
where it is more easily detectable.  
Higher quality spectra at low X-ray 
energies are needed to address this question.

It is interesting that QSOs do not show the same evidence for 
spectral flattening at energies $>10$ keV \cite{v98,c97}
that is common in Seyfert
galaxies\cite{np94} and is the signature of
Compton reflection.
Fe K$\alpha$ emission is also weaker and/or 
much less common in quasars and QSOs than in 
Seyfert galaxies and nearby radio-loud 
objects like 3C 120 \cite{g97}.
The equivalent width of 
the Fe K$\alpha$ line decreases with increasing 
luminosity from log $L_X = 41.7-47.2$ \cite{it95}.  This   
trend continues to z = 2 \cite{nandra97b} with  
RLQs at high z rarely showing 
Fe K$\alpha$ emission\cite{c97,reeves97}.

The lack of reflection and Fe K$\alpha$ emission 
in QSOs suggests that they possess a different structure in their 
accretion flow compared to lower-luminosity galaxies.   
The difference may relate to the high luminosities and 
high degree of ionization in the their inner regions.
A lack of Fe K$\alpha$ emission may signify complete ionization
of iron atoms in the accretion disk or that the inner parts   
of the disk have been completely blown away.  High ionization would 
also allow the Compton reflection to suffer much less
absorption in the disk.  In this case, reflection is 
still present but it would have a shape almost indistinguishable 
from the continuum source. 
For RLQs, the X-ray emission associated with 
the jet may simply dominate the spectrum,
rendering spectral features from the accretion flow undetectable.

\section*{Intrinsic X-ray Emission Mechanisms}

The source of the continuum emission in AGN is thought to be 
ultimately linked to an accreting supermassive black hole.  
Within the vicinity of the black hole, X-rays 
can be produced via Compton upscattering of soft photons  
off either electron-positron pairs\cite{zdz90},
a population of hot electrons\cite{tl95}, 
or bulk motion in the accretion 
flow\cite{tmk96,tmk97}.  The lack of annihilation lines
in Seyfert spectra causes some problems for pair 
models\cite{haardt97} and so this discussion focuses  
on the latter two mechanisms for X-ray production.
The emitting region is envisioned as
lying somewhere just above the accretion disk, often  
represented as a smooth extended corona above   
the disk or small-scale flaring regions on the disk surface.

Thermal Comptonization or bulk-motion Comptonization  
predict that the distinct observational signature
of a black hole is an emergent power law with a 
spectral turnover between $50-500$ keV.
If both processes are occurring, then the changes/differences
in photon index can be explained as a trade-off between 
the two.  If conditions
are such that bulk motion Comptonization dominates, then 
for an accretion disk that orbits a Schwarzschild black 
hole with an inner radius extending to the 
innermost stable orbit,
$\Gamma$ approaches 2.5 for mass accretion rates $m_{\odot} >> 1$,
where $m_{\odot} = {M_{\odot}}/{(M_{e})_{\odot}}$ and $(M_e)_{\odot}$
is the Eddington rate.  If thermal Comptonization dominates,
the spectra are harder, with $\Gamma \ge 1.5$.
This model has been successfully used to explain the spectra of   
Galactic black-hole candidates, which show 
$\Gamma$ = $1.3-1.9$ in their low (hard) 
state and $\Gamma$ = 2.5
in their high (soft) state \cite{ebisawa96}. 

For RQQs, where we think accretion mechanisms dominate
the X-ray spectrum, those 
with $\Gamma\ge2$ may be cases where we see
X-rays from bulk motion Comptonization without modifications
by ionized absorbers or reflection, i.e., the raw
emergent spectrum from the accreting black hole.
Within the context of the accretion model, the 
hard X-ray component in RQQs 
can result from thermal Comptonization.

Different mechanisms dominate the spectra of RLQs.  
RLQs show a significant correlation between their $0.1-2.4$
X-ray and total radio luminosity at 5 GHz \cite{b95}
and $\Gamma$ decreases with increasing radio loudness \cite{we87}
in the sense that flat spectrum radio quasars (FSRQs), which
are core-dominated,
have flatter X-ray spectra than their steep spectrum radio
quasar (SSRQ) counterparts\cite{bys97,wor87}.
This implies the presence of two emission mechanisms and 
has led to the two-component beaming model, which 
explains the difference in
observed properties as caused by the relative orientation of the
source axis to our line of sight \cite{bm87}.
In FSRQs, the highly beamed component points toward us
and we see mostly upscattering of low-energy photons off
relativistic electrons in the jet, while in
SSRQs we see mostly the isotropic emission. 

\section*{Unification and Evolution}

Unified schemes explain the diversity of
AGN classes as resulting from inclination plus obscuration effects,
e.g., Seyfert type 2 (narrow-line) galaxies are 
edge-on Seyfert 1 (broad-line) galaxies.
These models succeed when applied to specific  
subsamples of AGN, but a global unification scheme 
has proved elusive.
For radio-loud objects, two schemes seem to be required,
one for low luminosity objects and one for high luminosity
objects \cite{browne92}.  For RQQs, broad absorption line 
(BAL) QSOs can be unified 
with nonBAL QSOs through axis orientation,
but there are many arguments against RQQs being 
edge-on RLQs, including differences in their radio morphologies and 
underlying galaxy hosts \cite{barthel92}.  What we 
are left with is a fragmented ``unification'' scheme.

It may be more promising to examine evolution 
scenarios, i.e., are low luminosity Seyferts 
connected to distant and more powerful AGN? 
Seyfert galaxies have composite X-ray spectra that consist of an
underlying power law with $\Gamma\sim1.9$, 
a Compton reflection tail above $\sim10$ keV, Fe
K$\alpha$ emission\cite{np94}, and significant photoelectric
absorption from neutral and/or ionized gas\cite{r97}. 
High luminosity AGN lack significant evidence 
of X-ray reprocessing and many lack significant
absorption.   An evolutionary scenario that accounts 
for the difference requires 
QSOs to have much less cold material in their cores 
or for the material to be very highly ionized.
As the source becomes less luminous with time,
the ionization decreases and/or more cold material 
collects in the galaxy core. 

QSOs also differ from Seyferts by (apparently) having  
spectral components from two physically 
distinct emission mechanisms. 
One component is associated with a relativistic jet 
while the other is most likely related to the  
accretion mechanism.  Since processes associated 
with accretion are also thought to
produce the continuum emission 
in Seyfert galaxies, an evolutionary connection    
between RQQs and Seyferts is plausible.  
The soft X-ray spectra of RQQs are significantly steeper 
than Seyfert galaxies, but are also more  
consistent with the predictions of bulk-motion Comptonization.  
This would suggest that bulk motion Comptonization is 
the more important physical mechanism 
in RQQs while thermal Comptonization
of soft photons by a hot corona is the more 
important physical mechanism in Seyfert galaxies. 

\section*{The total energy output of the Universe}

Until recently, little attention has been given to sources of
energy in the universe that are not directly visible at optical-UV 
wavelengths.  It now seems probable that most AGN are
heavily absorbed, and that their central engines are primarily visible 
via hard X-rays. The energy density of the X-ray background peaks at 
$\sim30$ keV.  Less than $15\%$ of this total energy density can be accounted for
by the $Rosat$ AGN, which dominate the soft X-ray background.  If AGN
comprise the hard X-ray background, then most must have huge absorbing
columns (N$_{\rm H} \sim10^{22} - 10^{25}$ cm${-2}$)\cite{mgf94}.
The importance of the hard ($>10$ keV) X-ray band 
for studying the total energy output for these objects cannot 
be overemphasized.  
A significant fraction of the energy in the Universe may
reside in absorbed AGN \cite{fabian98b} and the total
accretion energy released by these AGN may be comparable to
the energy generated by nuclear burning by the total stellar 
population.  If most of the accretion in the 
Universe is highly obscured, then the emitted power per
galaxy based on optical, UV, or soft X-ray quasar luminosity functions 
will be underestimated.

\section*{Summary and Future Prospects}

This article summarizes our current knowledge of the 
X-ray spectral properties of QSOs. 
X-ray observations from 
z = 0.1 to $\sim3$ keV indicate two distinct X-ray 
components in their spectra.  One component is 
soft with $\Gamma\sim2.0-2.5$
and the other is hard with $\Gamma\sim1.5-1.9$. 
In the soft X-ray band the spectra flatten with increasing z  
while in the hard X-ray band the spectra show 
little change with z.  RLQs have flatter 
X-ray spectra than RQQs with the exception
of high energies at high z, where both have 
similar spectral shapes.  Accretion mechanisms such 
as thermal Comptonization or bulk-motion 
Comptonization are the most 
likely source of the continuum in RQQs while 
jet/beaming mechanisms dominate RLQs.
Significant absorption 
is observed in RLQs at high z but the physical properties of
the absorbing material are uncertain.   
More sensitive data will place stringent limits on 
the absorption cutoffs and properties of the absorbing 
gas, the spectral breaks between the soft and hard 
X-ray continuum components, and the 
signatures of X-ray reprocessing.  Considering an evolutionary 
scenario, RQQs and Seyfert galaxies are possibly
connected, with a different accretion mechanism 
dominating in each.

Results from X-ray experiments such as $ASCA$ for low-z 
AGN suggest that much  
of the accretion in the Universe is highly obscured.
So far, X-ray observations at intermediate to 
high z have shed little light on this question.
Surveys by $XMM$, ABRIXAS, and $Chandra$ (limited to $E<10$ keV)
will probe columns up to a few times 10$^{23}$
cm$^{-2}$; while future missions such as $Constellation-X$
(Valinia, this volume) will probe columns up to 10$^{25}$
cm$^{-2}$ for fluxes as low as
$\sim1\times10^{-14}$ ergs cm$^{-2}$ s$^{-1}$
in reasonable exposure times.

{
\begin{figure} 
\caption{Observed photon index in the soft X-ray band 
($0.1-2.4$ keV; $Rosat$) and hard X-ray band 
($\sim0.5-20$ keV; $Ginga$ and $ASCA$) vs. redshift.
}
\label{firstfigure}
\end{figure}

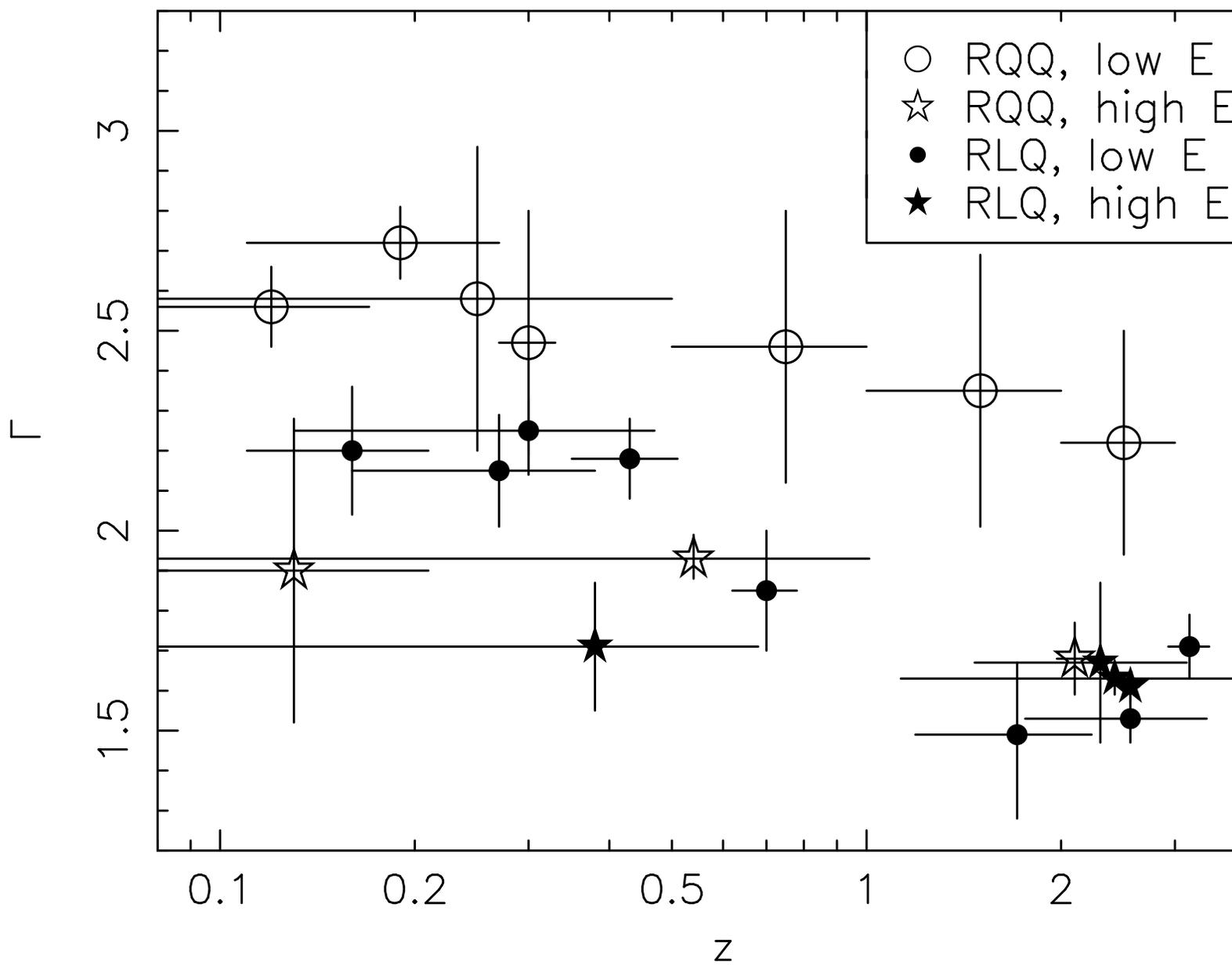
\end{document}